\documentstyle[12pt]{article}
\def\thf{\baselineskip=\normalbaselineskip\multiply\baselineskip
by 5\divide\baselineskip by 4}
\thf

\def\beq{\begin{equation}}
\def\eeq{\end{equation}}
\def\be{\beq} \def\fe{\eeq} \def\eqn{\label}

\def\spose#1{\hbox to 0pt{#1\hss}}\def\lta{\mathrel{\spose{\lower 3pt\hbox
{$\mathchar"218$}}\raise 2.0pt\hbox{$\mathchar"13C$}}}  \def\gta{\mathrel
{\spose{\lower 3pt\hbox{$\mathchar"218$}}\raise 2.0pt\hbox{$\mathchar"13E$}}} 
\def\Libra{\spose {--} {\cal L}}

\def\g{\gamma}  \def\calE{ {\cal E}}
  \def\kap{\kappa}\def\kap{\kappa}

\def\OmC{\Omega} \def\OmN{\Omega_{\rm n}}
\def\Lamb{\Lambda} \def\mN{m_{{\rm n}}}
\def\rhoC{\rho_{\rm c}} \def\rhoN{\rho_{\rm n}} 
\def\gamC{{\gamma}} \def\gamN{{\gamma_{\rm n}}} 
\def\uC{{u}} \def\uN{{u_{\rm n}}} 
\def\nB{{n_{\rm b}}} \def\nC{{n_{\rm c}}} \def\CC{\,{^\ast\!\nC}} 
\def\nN{{n_{\rm n}}} 
\def\aC{{a_{\rm c}}} \def\aNucl{{a_{\rm N}}} \def\NC{{N_{\rm c}}} 
\def\muC{\chi} \def\muN{\mu}   \def\vN{v_{\rm n}}
  
\def\Omplus{\Omega_{_+}} \def\Omminus{\Omega_{_-}} \def\Ompm{\Omega_{_\pm}}
\def\xiC{{\xi_{\rm c}}} \def\xiN{{\xi_{\rm n}}}
\def\fC{{f_{\rm c}}} \def\fN{{f_{\rm n}}}
 \def\fE{\overline f} \def\fA{\widehat f}
\def\TE{\overline T}  \def\TA{\widehat T}
\def\JE{\overline J} \def\calJE{\overline{\cal J}}
\def\JC{{J_{\rm c}}} \def\calJC{{{\cal J}_{\rm c}}}
\def\JN{{J_{\rm n}}} \def\calJN{{{\cal J}_{\rm n}}}
\def\calJh{\check{\cal J}}   \def\aN{{\alpha_{\rm n}}}

\def\fMagnus{f_{_{\rm lift}}} \def\FMagnus{F_{_{\rm lift}}}
\def\fdrag{f_{_{\rm drag}}}       \def\fNdrag{\fN_{\!_{\perp}}}  
\def\Fdrag{F_{_{\rm drag}}}  \def\Ftension{F_{_{\rm tens}}}

\def\calF{{\cal F}} 

\def\Rcurv{R_{_{\rm c}}}  

 \def\drag{{\cal C_{\rm r}}} \def\drac{{c_{\rm r}}}  
\def\csound{{c_{\rm s}}}
\def\chemr{\Xi\,} 
 \def\delp{\delta_{_\Vert}}

    \def\calR{{ \cal R} }    \def\Si{{\mit \Sigma}}
\def\kt{k} \def\mh{h} \def\ph{h_{_\perp}}

\def\nabl{\nabla\!}
\def\ww{w}

\def\kFerme{k_{\rm e}}
\def\kFermi{k_{_{\rm n}}} \def\EFermi{E_{_{\rm n}}}
\def\xiP{{\xi_{\rm p}}} 
\def\xiN{{\xi_{\rm n}}} \def\DelP{{\Delta_{_{\rm P}}}}
\def\bPin{{b_{_{\rm P}}}}  \def\EPin{{E_{_{\rm P}}}}  
\def\FPin{{F_{_{\rm P}}}}
\def\mP{{m_{\rm p}}} \def\DelmP{\Delta\mP}
\def\entra{\alpha} \def\penetr{{\delta_{\rm p}}}

\begin{document}

\begin{center}
{\bf Differential rotation of relativistic superfluid in neutron stars.}\\
\vskip 2cm
{\bf David Langlois} \\
{\it D\'epartement d'Astrophysique Relativiste et de Cosmologie,\\
Centre National de la Recherche Scientifique, \\Observatoire de
Paris, 92195 Meudon, France}\\
\vskip 1cm
{\bf David M. Sedrakian}\\
{\it Erevan State University, Erevan, Republic of Armenia}\\
{\it and}\\
{\it D\'epartement d'Astrophysique Relativiste et de Cosmologie,\\
Centre National de la Recherche Scientifique, \\Observatoire de
Paris, 92195 Meudon, France}\\
\vskip 1cm
{\bf Brandon Carter} \\
{\it D\'epartement d'Astrophysique Relativiste et de Cosmologie,\\
Centre National de la Recherche Scientifique, \\Observatoire de
Paris, 92195 Meudon, France}\\
\vskip 1cm
{2 May 1997}\\
\end{center} 

{\bf Abstract} It is shown how to set up a mathematically elegant and fully
relativistic superfluid model that can provide a realistic approximation
(neglecting small anisotropies due to crust solidity, magnetic fields, et
cetera, but allowing for the regions with vortex pinning) of the global
structure of a rotating neutron star,  in terms of just two independently
moving constituents, one of which represents the differentially rotating
neutron superfluid, while the other part represents the combination of all the
other ingredients, including the degenerate electrons, the superfluid protons
in the core, and the ions in the crust, whose electromagnetic interactions
will tend to keep them locked together in a state of approximately rigid
rotation. Order of magnitude estimates are provided for relevant parameters
such as the resistive drag coefficient and the maximum pinning force.
\vfill
\eject

\section{Introduction}
\label{Sec1}

A considerable body of observational information about neutron
behaviour under various circumstances is now available from pulsar
timing measurements. It is generally recognised that many features can
be understood only if it is assumed that -- as predicted on theoretical
grounds -- a substantial part of the interior of such a star is in a
superfluid state.  It is evident that the observations should be
interpreted as providing a direct measurement of the angular velocity,
$\OmC$ say, of the solid outer crust of the star, with respect to which
the magnetosphere responsible for the pulsed emission can be presumed
to be rigidly corotating. However due to the superfluidity the coupling
of the crust to the neutron fluid interior may be very weak, so that
the latter will have a locally variable angular velocity, $\OmN$, say
that may differ very significantly from the angular velocity $\OmC$ of
the rigidly rotating exterior.

\par In typical cases the outer part will be steadily slowing down, so
that with the sign convention that $\OmC$ itself should be positive,
$\OmC>0$, one will observe a negative value $\dot\OmC<0$ for the rate
of change due to the angular momentum loss involved in the pulsar
radiation process. In such circumstances one expects that the
corresponding slow down of the weakly coupled superfluid interior will
be subject to a delay, so that it will be in a state of relatively
rapid rotation with $\OmN>\OmC$. However there will also be less usual
circumstances in which this inequality might be reversed, $\OmN<\OmC$,
for example during a period of spin up with $\dot\OmC>0$ due to
accretion.  (It is even possible to conceive in which accretion is
absent, and in which the inner part is still in a state of relatively
rapidly rotation, $\OmN>\OmC$, while the outer part is gradually
spinning up, $\dot\OmC>0$ due to the weak transfer of angular momentum
from the interior, which may eventually become more imporant than the
effect of pulsar radiation drag if the magnetosphere finally gets
aligned with the rotation axis.)

\par The present work is concerned with  quantitative evaluation of
such effects, primarily as a contribution to the understanding of 
the long term evolution of the star. However this work will also
be relevant to the more spectacular short term events known as
``glitches'', namely sudden angular velocity increases, of which the
largest are characterised by ${\Delta\OmC}/{\OmC} \sim 10^{-6}$, that are
followed by a period of continuous relaxation,  as well as so 
called ``noises" (fluctuations with $|\Delta\OmC|/\OmC\sim
10^{-9}$) \cite{mnhg}-\cite{flanagan93}.  Such effects provided the
main motivation for  much of the theoretical attention that has been
directed, in the last two decades, to the dynamics of the neutron
superfluid in the neutron stars \cite{pines}\cite{SSh91}.

\par All these irregularities of the angular velocity  are superimposed
on the long term ``secular'' variation, and are small in comparison
with the absolute angular velocity of the star, but they are very
significant for the  understanding of the physics of pulsar interior.
If the electromagnetic  nature of the secular variation is reasonably
well understood, the basic physical mechanism responsible for the
 rotational irregularities is still a matter of scientific debate. One
of the most important but still controversial aspects concerns the
``pinning'' effect, whereby the vortex lines associated with the
rotation of the superfluid are more or less strongly attached to the
ionic lattice forming the solid crust, whose lower layers (at densities
of about $10^{11}$ gm/cm$^3$ and upwards) are interpenetrated by the
neutron superfluid.

\par In the pinned regime, when the superfluid velocity relaxes by
means of vortex creep, the theoretical superfluid relaxation times are
found to be compatible with the observed postjump relaxation timecales
\cite{aaps}\cite{leb}. If the pinning or localization of neutron
vortices is not effective, they interact relatively weakly with the
electron-phonon system either directly \cite{Jones90} \cite{Jones92} or
through the excitation of the oscillatory degrees of freedom of vortex
lines by the vortex-nucleus interaction \cite{eb}. Whether the pinned
or the free flow state is operative in the inner crust depends on
several uncertain factors like relative orientation of nuclear and
neutron vortex lattices, the strength of the pinning potential, the
timescale for repinning, etc.

\par An alternative model\cite{SS95a} for the dynamical coupling of the
neutron star superfluid core is based on the dynamics of a neutron vortex
with strong magnetic field. Because of the strong dependence of the
vortex flow viscosity coefficient on the  matter density, the core
superfluid has a wide range of dynamical coupling times, which are
consistent with the observed postjump relaxation time constants
\cite{cdk}. This provides the basis for a theory of nonstationary
dynamics of neutron star core rotation \cite{ssct}\cite{SS95b} that --
when  applied to the analysis of angular velocity jumps and postjump
relaxations -- can explain the observational data for the first 6
glitches of Vela pulsar \cite{cdk}, and also gives correct values of
the mean dynamical time (glitch or postglitch relaxation) for other
pulsars in which glitches have been observed\cite{ssct}.

\par In so far as allowance for superfluidity is concerned, all the work
mentionned above was carried out within the framework of a Newtonian
framework, for which a for detailed phenomenological treatment has by now been
developed\cite{SSh91}\cite{Men91}\cite{MenLin91}. However it is well known that relativistic
corrections to the Newtonian theory for neutron stars are of the order of
20-30$\%$ , and so may be important in relation to the effects of rotation,
which are of the same order. This is the reason why it has long been standard
practice\cite{BGSM93}\cite{SBGH94} for numerical work on the basic structure of neutron stars
-- for which a simple perfect fluid description suffices -- to use a fully
relativistic treatment. It is only due to technical difficulties that it has
not yet become standard practice to use a similarly relativistic treatment for
the study of more detailed effects for which a simple perfect fluid
description is insufficient.

\par Although the essential theoretical machinery needed for a
similarly relativistic treatment of the effects of the solidity of the
crust was made available quite a long time
ago\cite{CQ72}\cite{C73}\cite{CQ75}, the technical complications
involved in actually applying this machinery are such that its
effective exploitation has only recently begun to be feasible in
practice\cite{Priou92}. In so far as the effects of superfluidity are
concerned, the situation was somewhat different, since the machinery
needed for a fully relativistic treatment was not available at all. The
standard formalism of irrotational perfect fluid mechanics would
suffice in the zero temperature limit if no vortex lines were
present\cite{KY95} but  this simplification is unjustifiable unless
superfluid angular velocity $\OmN$ is infinitesimally small compared
with what typically occurs. In a realistic description the superfluid
will be effectively fibrated by a dense latice of quantised vortex
lines. Although a Newtonian description was already
available\cite{SSh91}\cite{Men91}\cite{MenLin91}, what  was lacking
for a relativistic description was an appropriate way of allowing at a
macroscopic level for the local anisotropy due to the microstructure
formed by the quantised vortex lines in the rotating superfluid
interior. An elegant variational model of the kind required for this
purpose has however been recently developped\cite{CL95c}. It
fortunately turns out that the actual implementation of the relevant
vortex fibration machinery is not quite as complicated as that of the
elastic solid machinery\cite{Priou92} needed for treating elastic
deformations of the crust.

\par The models introduced so far for the relativistic treatment of the
isotropies due both to elastic solidity in the crust\cite{CQ72} and to
superfluid vorticity in the interior\cite{CL95c} are all subject to the
limitation that the solid or superfluid involved is supposed to be
strictly conserved.  Although it is easy to allow for deviations from a
strictly conservative behaviour by allowing for resistive (thermal,
electric, or more general) conductivity\cite{C89} it is not so obvious
how these rather elaborate models should be modified to allow for
what we refer to as ``transfusion'', meaning the
transfer of matter (by the ``neutron drip'' process at densities above
about $10^{11}$ gm/cm$^3$) from the solid ionic lattice to the ambient
superfluid, a process that will usually occur too slowly to be
important on short timescales, but that will be significant in the long
term readjustment of the stellar equilibrium in order to allow for the
effect of substantial angular momentum loss. Since the deviations from
local isotropy due to elastic solidity or superfluid vorticity are
never expected to exceed a small fraction of a percent, and those due
to magnetic fields will also be fairly small, they can reasonably be
neglected as a first approximation for the purposes of describing the
long term evolution of the star. It will be shown below that within the
framework of such an approximation, i.e. subject to the postulate that there
is no intrinsic anisotropy, it is easy to set up a new kind two-fluid model
that can adequately describe the transfusion process whereby matter is
transferred between the crust and the neutron superfluid in a manner that will
be adequate for our present purpose and that is likely to be useful for future
applications.

\par The application for which this innovation is intended in the present work
is a step in the bridging of the gap between previous work that used a
fully relativistic treatment but that was based on the approximation of a
perfect fluid description, and previous work that was based on a more refined
description of the neutron star matter, but that used a Newtonian
approximation for the large scale geometry. The present investigation does not
include the more refined modifications, allowing for effects such as proton
superconductivity, that have already been investigated in the Newtonian
approximation\cite{SSh91}\cite{Men91}\cite{MenLin91}, but 
whose relativistic description  is left for future work.

\par The approach followed here will be rigourously theoretical in the
sense that we shall refrain from making ad hoc parameter
adjustments in order to match observational results
whose correct interpretation may still be open to question.

\section{Transfusive two constituent superfluid model}
\label{Sec2}

\subsection{General principles.}
\label{2_1}

Unlike the (non-transfusive) Landau type two constituent superfluid
model that has been developped in recent
years\cite{LK82}\cite{CK92}\cite{CL95a} to provide a microscopic (inter
vortex) description allowing for the independent entropy current that
will be present in a relativistic superfluid at finite temperature, the
essentially different kind of two constituent superfluid model set up
here allows for what we shall refer to as {\it transfusion}, meaning
the transfer of material between the the distinct constituents which
are not separately conserved.  In a transfusive model of the type set
up here, the ``normal'' constituent is not entirely dependent on
(though it does include) entropy, so that it is present even at zero
temperature: the primary role of this non-superfluid constituent is to
represent the fraction of the baryonic material of the neutron star
that is not included in the neutron superfluid, as well as the
degenerate electron gas that will be present to neutralise the charge
density resulting from the fact that some of these baryons will have
the form of protons rather than neutrons. In the solid ``crust'' layers
of a neutron star the protons will be concentrated together with a
certain fraction of the neutrons in discrete nuclear type ions, which
at the relatively moderate temperatures that are expected to apply will
form a solid lattice. In the upper crust the ``normal'' constituent
consisting of the ionic lattice and the degenerate electrons will
include everything, but in the lower crust (at densities above about
$10^{11}$ gm/cm$^3$) the crust will be interpenetrated by an independently
moving neutron superfluid. What we refer to as ``transfusion'' occurs
when compression takes place so that the ionic constituent undergoes a
fusion process whereby neutrons are released in the form of newly
created superfluid matter, or conversely, when relaxation of the
pressure allows excess neutrons to be reabsorbed into the ions.

A more elaborate treatment would specifically allow for the expectation that
the protons would form an independently conducting superfluid of their own at
very high densities, whereas they will combine with some of the neutrons at
intermediate densities, and with all of the neutrons at low densities, to form
discrete ions which will tend to crystalise to form a possibly anisotropic
lattice. What matters for our present purpose is that regardless of its
detailed constitution, all this ``normal'' matter will in effect be strongly
self coupled\cite{ALS84} by short range electromagnetic interactions so that it movement
will be describable to a very good approximation as that of a single fluid
with a well defined 4-velocity, $\uC{^\mu}$ say, the only independent motion
being that of the (electromagnetically neutral) neutron superfluid with
velocity $\uN{^\mu}$ say.  The latter will specify the direction of the part
of the baryon current 
 \be \nN{^\mu}=\nN \uN{^\mu}\eqn{2.1}\fe 
carried by the neutron superfluid, while the ``normal'' matter velocity
specifies the direction of the remaining {\it  collectively comoving} part 
 \be \nC{^\mu}=\nC\uC{^\mu} \eqn{2.2}\fe 
of the baryon current. 

Under conditions of stationary circular flow round the axis of symmetry of the
star, each of these currents will be separately conserved (making it feasible
to use the more elaborate non-anisotropic models that are available
\cite{CQ72}\cite{CL95c}). However, during active phases of the stellar life,
a certain amount of {\it interchange} of matter may take place between the two
constituents due to the occurrence of {\it convection}: to be more explicit,
there may be regions of rising and descending flow within which baryons are
transferred respectively from or to the superfluid, so that only the total
baryon current
 \be \nB{^\mu}=\nN{^\mu}+\nC{^\mu} \eqn{2.3}\fe
is conserved,
\be \nabl_\mu\nB{^\mu}=0\, ,\eqn{2.4}\fe
while the separate divergence contributions $\nabl_\mu\nN{^\mu}$ and
$\nabl_\mu\nC{^\mu}$ can be non-zero. 

At densities below the ``neutron drip'' transition at about $10^{11}$
gm/cm$^3$, the ``normal''  collectively comoving constituent
$\nC{^\mu}$ will of course be identifiable with the total,
$\nB{^\mu}$.  The reason why the remaining free neutron part
$\nN{^\mu}$ -- which will always  be present at higher densities -- is
presumed to be in a state of superfluidity is that the relevant
condensation temperature, below which the neutrons form bosonic
condensate of Cooper type pairs is estimated \cite{E88} to be at least
of the order of $10^9$ K, while it is expected that a newly formed
neutron star will drop substantially below this temperature within a
few hundred months\cite{Ts79}. At such comparatively low temperatures
the corresponding entropy current $s^\mu$ say will not play a very
important dynamical role, but for the sake of exact internal
consistence it will be allowed for in the model set up here, in which
it will be taken for granted that it forms part of the ``normal''
collectively comoving constituent so that it will have the form
 \be s^\mu= s\uC^\mu\, .\eqn{2.5}\fe

Under conditions of sufficiently slow convection, the transfer needs
not involve significant dissipation, so  the process should be
describable by a Lagrangian scalar, $\Lamb$ say, that will depend just
on the currents introduced above, of which the independent components
are given just by the vectors $\nC^\mu$ and $\nN^\mu$ and the scalar
$s$. Except at the highest densities,  at which the distinct ions cease
to exist, it would probably be a good approximation to suppose that the
Lagrangian separates in the form $\Lamb=-\rhoC-\rhoN$ in which $\rhoC$
is an energy density depending only on $s$ and $\nC$, while $\rhoN$ is
an another energy density depending only on $\nN$, but we shall not
invoke such a postulate here, i.e. we allow for the likelihood that,
particularly at high densities, beyond about $10^{13}$ gm/cm$^3$, the
properties of ``normal'' constituent will be affected by the presence
of the superfluid constituent and vice versa, which means that there
will be an {\it entrainment} effect\cite{AB76}\cite{SSh91}
\cite{ALS84}\cite{Sj76},
whereby for example the velocity of the superfluid neutron current will
no longer be parallel to the corresponding momentum. (As an alternative
to the more suitable term ``entrainment'' this mechanism is sometimes
referred to in the litterature as ``drag'', which is misleading because
entrainment is a purely conservative, entirely non-dissipative effect,
whereas the usual kinds of drag in physics, and in particular the kind
of drag to be discussed below, are essentially dissipative processes.)

If we adoped the (gas type) description embodied in the separation
ansatz we would have two separate variation laws which in a fixed
background would take the form $\delta\rhoC=\Theta \delta s+\muC
\delta\nC$ and $\delta \rhoN = \muN \delta\nN$, in which $\Theta$ would
be interpretable as the temperature, $\muC$ would be interpretable as
the relativistic chemical potential per baryon in the ``normal'' part,
and $\muN$ would be interpretable as the relativistic chemical
potential per baryon (in other words the effective mass per neutron) in
the superfluid part (which would be equal to its analogue in the
``normal'' part, i.e. $\muN=\muC$, in the particular case of a state of
static thermodynamic equilibrium.)

In the less specialised (liquid type) description to be used here, there will
just be a single ``conglomerated'' variation law, whose most general form,
including allowance for a conceivable variation of the background metric, will
be expressible as
 \be \delta\Lamb=-\Theta\delta s +\muC{_\nu} \delta\nC^\mu+ 
 \muN{_\nu}\delta\nN^\nu +{_1\over^2}\big(\nC{^\mu}\muC{^\nu}
 +\nN{^\mu}\muN{^\mu}\big)\delta g_{\mu\nu}\, ,\eqn{2.6}\fe
where $\Theta$ is to be interpreted as the temperature and where $\muN_\mu$
and $\muC_\mu$ are to be interpreted as the 4-momentum per baryon of the
neutron superfluid and the ``normal'' constituent respectively.

To obtain suitable fluid type dynamical equations from a Lagrangian expressed
as above just in terms of the relevant currents, the variation of the
latter must be appropriately constrained in the manner\cite{C89} that was
originally introduced for the case of a simple perfect fluid by Taub. The
standard Taub procedure can be characterised as the requirement that the
variation of the relevant current three form, which for the ``normal''
constituent in the present application will be 
 \be \CC_{\mu\nu\rho}=\varepsilon_{\mu\nu\rho\sigma}\nC^\sigma\, ,\eqn{2.7}\fe
should be given by Lie transportation with respect to an associated, freely
chosen, displacement vector field $\xiC{^\mu}$ say. This ansatz gives the well
known result 
 \be \delta \CC_{\mu\nu\rho}=\xiC^\lambda\nabl_{\lambda} \CC_{\mu\nu\rho}
 +3\CC_{\lambda[\mu\nu}\nabl_{\rho]}\xiC{^\lambda}\ .\eqn{2.8}\fe
Although a variation $\delta g_{\mu\nu}$ of the metric has no effect on the
fundamental current three form, $\CC_{\mu\nu\rho}$, it will contribute to the
variation of the corresponding vector,
 \be \nC^\mu={1\over 3!}\varepsilon^{\mu\nu\rho\sigma}\CC_{\nu\rho\sigma}\, , 
 \eqn{2.9}\fe
for which one obtains
 \be \delta \nC^\mu=\xiC^\nu\nabl_\nu\nC^\mu-\nC^\nu\nabl_\nu\xiC^\mu
 +\nC^\mu\big(\nabl_\nu\xiC^\nu-{_1\over^2}\g^{\nu\rho}\delta g_{\nu\rho}
 \big) \eqn{2.10}\fe  
in terms of the orthogonally projected metric,
\be \gamC{_{\mu\nu}}=g_{\mu\nu}+\uC_\mu \uC_\nu. \eqn{2.11}\fe
The corresponding variation of the unit flow vector will be given by
 \be \delta \uC{^\mu}=\gamC{^\mu}_{\ \rho}\big(\xiC^\nu\nabl_\nu\uC^\rho-
 \uC{^\nu}\nabl_\nu\xiC{^\rho}\big)-{_1\over^2}\uC^\mu\uC^\nu\uC^\rho
 \delta g_{\nu\rho}\, ,\eqn{2.12}\fe
and the corresponding variation in the current amplitude $\nC$ will be
 \be \delta \nC=\nabl_\nu\big(\nC\xiC^\nu\big)
 +\nC\big(\uC{^\mu}\uC{^\nu}\nabl_\mu\xiC{_\nu}
 -{_1\over^2}\gamC{^{\mu\nu}}\delta g_{\mu\nu}\big)\, .\eqn{2.13}\fe
Since the entropy flux is to be considered as comoving with the ``normal''
constituent, it is subject to a variation given by the same displacement
vector $\xiC$, which thus gives 
 \be \delta s=\nabl_\nu\big(s\xiC{^\nu}\big)+s\big(
 \uC{^\mu}\uC{^\nu}\nabl_\mu\xiC{_\nu}
 -{_1\over^2}\gamC^{\mu\nu}\delta g_{\mu\nu}\big)\, .\eqn{2.14}\fe
On the other hand for the superfluid constituent there will be an
independent displacement vector field $\xiN^\mu $ say,
in terms of which the analogously constructed variation will be
 \be \delta \nN^\mu=\xiN^\nu\nabl_\nu\nN^\mu-\nN^\nu\nabl_\nu\xiN^\mu
 +\nN^\mu\big(\nabl_\nu\xiN^\nu-{_1\over^2}\g^{\nu\rho}\delta g_{\nu\rho}
 \big)\, .\eqn{2.15}\fe  

The effect of this variation process on the Lagrangian density
$\Vert g\Vert^{1/2}\Lamb$ itself can be seen to be expressible
in the standard form
 \be \Vert g\Vert^{-1/2}\delta\big(\Vert g\Vert^{1/2}\Lamb\big)=
 \xiC{^\nu}\fC{_\nu}+\xiN{^\nu}\fN{_\nu}+{_1\over^2}\TE{^{\mu\nu}}
 \delta g_{\mu\nu}+\nabl_\mu\calR^\mu\, ,\eqn{2.16}\fe
in which $\fC{_\nu}$ will be interpretable as the force density acting on the
``normal'' constituent, $\fN{_\nu}$ will be interpretable as the force density
acting on the superfluid constituent, $\TE{^{\mu\nu}}$ will be interpretable as
the stress momentum energy density of the two constituent as a whole. The
residual current $\calR^\mu$ in the divergence will be of no importance for
our present purpose (by Green's theorem it just gives a surface contribution
that will vanish by the variational boundary conditions) but it is to be noted
for the record that it will have the form
 \be \calR^\mu= 2\xiC{^{[\mu}}\uC{^{\nu]}} \big(s\Theta\uC{_\nu} + 
 \nC\muC_\nu\big) + 2\xiN\,{^{[\mu}}\nN{^{\nu]}}\muN{_\nu}\, .\eqn{2.17}\fe
The conglomerated stress momentum energy density tensor can easily be read 
out as
 \be \TE{^\mu}_{\nu}=\Psi g^\mu_{\ \nu}+s\Theta\uC{^\mu}\uC{_\nu}
 +\nC{^\mu}\muC{_\nu} + \nN{^\mu}\muN{_\nu} \eqn{2.18}\fe
where
 \be \Psi=\Lamb+s\Theta-\nC{^\nu}\muC{_\nu}-\nN{^\nu}\muN{_\nu}\ .\eqn{2.19}\fe
(Although this expression is not manifestly symmetric, the asymmetric
contributions will automatically cancel due to the identity
$\nC{^{[\mu}}\muC{^{\nu]}}=\muN{^{[\mu}}\nN{^{\nu]}}$).
What matters most for our present purpose is the form of the respective
force densities: the force law (i.e. the relevant relativistic 
generalisation of Newton's ``second'' law of motion) for
the ``normal'' constituent is found to take the form
 \be \fC{_\nu}= 2s^\mu\nabl_{[\mu}\big(\Theta \uC_{\nu]}\big)
 +2\nC{^\mu}\nabl_{[\mu}\muC{_{\nu]}}+\Theta \uC{_\nu}\nabl_\mu s^\mu
 +\muC{_\nu}\nabl_\mu \nC^\mu\, ,\eqn{2.20}\fe
while the force law for the superfluid component is found to take
the simpler form
 \be \fN{_\nu}=\nN{^\mu}\ww_{\mu\nu}
 +\muN{_\nu}\nabl_\mu \nN^\mu\, ,\eqn{2.21}\fe
using the notation
 \be \ww_{\mu\nu}=2\nabl_{[\mu}\muN{_{\nu]}}\eqn{2.22}\fe
for the vorticity 2-form of the superfluid.

\subsection{Non-dissipative ``free'' and ``pinned'' limit models.}
\label{2_2}

Up to this point we have been dealing only with purely
kinematical relationships, without making any physical assumptions
about the form of the dynamical equations of motion. If we were
to postulate that the latter were given simply by application of the
variation principle for freely chosen displacement fields
$\xiC{^\mu}$ and $\xiN{^\mu}$ we would obtain dynamical equations
given just by the condition that the force densities $\fC{_\nu}$
and $\fN{_\nu}$ should each vanish separately, a postulate that is
too restrictive for our present purpose since it would entail the
separate conservation of $\nC{^\mu}$ and $\nN{^\mu}$, which
would not be realistic for scenarios involving convection.

Before describing the more appropriate force ansatz that will be adopted
below, it is to be remarked that the forces can not be specified in an
entirely independent manner, in view of the action-reaction identity (the
relativistic generalisation of Newton's ``third'' law) that is derivable from
the consideration that the system will evidently be globally unaffected if
both currents undergo {\it the same} displacement, $\xiN{^\mu}=\xiC{^\mu}$
provided that the metric itself is subject to the corresponding gauge
adjustment, namely $\delta g_{\mu\nu}=2\nabl_{(\mu}\xiC_{\nu)}$. It can be seen
from  the basic variation identity (\ref{2.14}) that since the action must be
invariant with respect to any variation of this trivial kind (which merely
represents an infinitesimal coordinate transformation) the forces must be
subject to an identity of the form
 \be \fC{_\nu}+\fN{_\nu}=\fE{_\nu} \eqn{2.23}\fe
where $\fE{_\nu}$ is the conglomerated {\it external force density}
that is defined by
 \be \fE{_\nu}=\nabl_\mu \TE{^\mu}_\nu \ .\eqn{2.24}\fe
For a system that is isolated in the strictest sense the external force
density would simply vanish, but for the application to be considered here it
will be necessary to take account of the action of a {\it non zero} external
force density $\fE{_\mu}$ on the star, in order to allow for the backreaction
(and in particular the angular momentum loss) due to radiation by the outer
magnetosphere.

Before including allowance for this and other potentially dissipative effects,
it is worthwhile to present the simplest relevant model, in which interchange
is appropriately allowed for in a conservative manner so that one has
$\nabl_\mu s^\mu=0$ even though $\nabl_\mu\nC{^\mu}= -\nabl_\mu\nN{^\mu}\neq
0$. The obvious way to obtain the requisite model within the present framework
is as follows. To start with it is of course necessary in this particular case
to suppose that there is no external force acting on the star, i.e.
\be \fE{_\mu}=0\, .\eqn{2.25}\fe
Whereas the preceeding equations in this section have been kinematic
identities of a mathematically obligatory nature, (\ref{2.25}) is the first
case of a physical assumption of the kind that can be, and later on will be,
relaxed in a more general treatment. The assumption (\ref{2.25}) evidently
entails by (\ref{2.23}) that the force densities acting between the two
constituents will have to be equal and opposite, i.e. 
\be \fN{_\mu}=-\fC{_\mu}\, ,\eqn{2.26}\fe
so it suffices to choose the physical prescription for just one of them in
order to specify the other and thus to fully determine the dynamical
evolution. Although the complete expression (\ref{2.20}) is not so simple, it
is to be observed that the time component in the ``normal'' rest frame
(representing the rate of working on the ``normal'' constituent) as obtained
by contraction with the relevant unit vector $\uC{^\nu}$ has the comparitively
simple form
 \be \uC{^\nu}\fC{_\nu}=\uC{^\nu}\muC{_\nu}\nabl_\mu\nC{^\mu}-\Theta
 \nabl_\mu s^\mu\, .\eqn{2.27}\fe
In view of the total baryon current conservation law (\ref{2.3}) this
will be consistent with entropy conservation
 \be \nabl_\mu s^\mu=0  \eqn{2.28}\fe
if and only if the ansatz for the force $\fN{_\mu}$ acting on the
superconducting constituent is such that the condition
 \be \uC{^\nu}\fN{_\nu}=\uC{^\nu}\muC{_\nu}\nabl_\mu \nN{^\mu}
 \eqn{2.29}\fe 
is satisfied. This requirement is expressible in the form
 \be \nN{^\mu}\ww_{\mu\nu}\uC{^\nu}=\uC{^\nu}(\muC{_\nu}-
 \muN{_\nu})\nabl_\mu \nN^\mu\, ,\eqn{2.29a}\fe 
in which the right hand side would obviously vanish for a non-transfusive 
model, as characterised by the separate conservation law
 \be \nabl_\mu \nN^\mu=0\ .\eqn{2.31a}\fe 
The right hand side of (\ref{2.29a}) will also vanish for a transfusive
model of the kind more relevant to the neutron star applications under
consideration here, in which -- except for phenomena with timescales so
very short as to be comparable with those of the weak interactions
involved in the creation of protons from neutrons -- it can be taken as
a very good approximation that the condition
 \be \uC{^\nu}(\muC{_\nu}- \muN{_\nu})=0\, ,\eqn{2.31b}\fe 
expressing transfusive (``chemical'' type) equilibrium between the
superfluid and ``normal'' constituents with respect to the ``normal''
rest frame, will be satisfied instead.

Whichever of the alternatives (\ref{2.31a}) and (\ref{2.31b}) is used,
the requirement that the complete system of equations of motion be
consistent with the entropy conservation condition (\ref{2.28}) will
simply reduce to the condition that the remaining dynamical equations
should be such as to ensure that left hand side of (\ref{2.29a}) will
vanish.  There are just two obvious ways of achieving this
requirement.  The first way is to postulate that the superfluid
momentum transport equations should be formally the same as in an
ordinary single constituent barotropic perfect fluid, meaning that they
should be given by the familiar expression
 \be \nN{^\mu}\ww_{\mu\nu}=0 \, . \eqn{2.30a}\fe 
This ``free limit'' equation of motion is interpretable as the condition of
conservation of the superfluid vorticity flux across any two dimensional
surface that is comoving with the superfluid current $\nN{^\mu}$. The other
way is to postulate that the superfluid dynamical equations have the
alternative form
 \be \uC{^\mu}\ww_{\mu\nu}=0 \, , \eqn{2.30b}\fe which is interpretable
as the condition of conservation of the superfluid vorticity flux across any
two dimensional surface that is comoving not with respect to the superfluid
but with respect to the other ``normal'' constituent current $\nC{^\mu}$.  The
latter variant is the equation of motion that is appropriate in regions where
vortex pinning is effective. It will be seen from the order of magnitude
estimates provided in Subsection\ref{6_2} that this ``pinned limit'' model,
i.e. the model based on the use of (\ref{2.31b}) in conjuction with
(\ref{2.30b}), will also provide a very good approximation in the deep core
region of the star where that resistive drag by the ``normal'' constituent
turns out to be extremely large, so that although they are not pinned in the
strictest sense the vortices will in effect be almost exactly comoving with
the ``normal'' constituent.  On the other hand it will be seen that the first
of these possibilities, i.e.  the ``free limit'' model based on the use of
(\ref{2.31b}) in conjuction with (\ref{2.30a}), should provide a very good
approximation in much of the lower crust region where that resistive drag
exerted on the vortices by the ``normal'' constituent turns out to be
extremely small.

\subsection{Dissipative interactions.}
\label{2_3}

Our purpose in the present subsection is to  set up a more general model that
interpolates between the non-dissipative ``free'' and ``pinned'' extremes
presented in the preceeding subsection, so as to allow for dissipative
interaction between the two constituents, the kind that is most important for
the application discussed below being resistive drag. For this purpose we
shall retain only the general framework of  Subsection \ref{2_1} but not the
more specialised conditions described in Subsection \ref{2_2}.  With reference
to the latter, the only stage for which the superfluidity property of the
neutrons is directly relevant is the vorticity transport law, for which,
instead of the idealised extreme alternatives (\ref{2.30a}) and (\ref{2.30b})
we need an intermediate modification to allow for the finite resistive drag
force\cite{Jones90} exerted by the flux of the normal constituent on the cores
of the microscopic vortices.

Since in the present work we are neglecting the energy and tension of
the vortices (whose -- relatively small -- dynamic effects have
recently been the subject of analysis in their own right \cite{CL95c}),
the form of the vorticity transport law in the conservative limit
governed by (\ref{2.30a}) is, as remarked above, just the same as it
would be for a constituent of the ordinary perfectly fluid but not
superfluid type. The feature that the superfluidity property as such
does not have any ostensible role in this model contrasts with the
situation that would apply in a microscopic description, for which the
relevant vorticity 2-form $\ww_{\mu\nu}$ would vanish. It also
contrasts with the situation that applies when drag between the normal
and superfluid constituents needs to be taken into account.

The way in which the superfluidity property remains important at a macroscopic
level (even when the resulting anisotropy\cite{CL95c} is neglected) can be
explained as follows. If the neutron fluid were of the ordinary non superfluid
kind, the drag force contribution $\fNdrag{^\mu}$ say, would be
aligned with (and roughly proportional to) the relative flow vector $\uC^\mu+
\uC{^\nu}\uN{_\nu}\uN{^\mu}$. However a resistivity force of this familiar
kind is not compatible with the property of superfluidity, which does not only
require that the (macroscopically averaged) vorticity  should be represented
by a closed 2-form $\ww_{\mu\nu}$ as in conservative and dissipative fluid
models of a more general kind: it is also necessary in the superfluid case
that the vorticity 2-form should be consistent with a microscopic description in
which the vortex cores are localised on 2-dimensional string type world
sheets, which means that it should satisfy the algebraic degeneracy condition
\be \varepsilon^{\mu\nu\rho\sigma}\ww_{\mu\nu}\ww_{\rho\sigma}=0\, 
.\eqn{3.1}\fe
In conjunction with the relevant integrability condition, which is just
the  closure property $\nabl_{[\mu}\ww_{\nu\rho]}=0$ that results
automatically from the construction of the vorticity according to (\ref{2.22})
as the exterior derivative of a momentum form, the degeneracy condition
(\ref{3.1}) ensures the existence of a congruence of two dimensional
worldsheets orthogonal to the vorticity 2-form, which will be expressible
in terms of its scalar amplitude, 
 \be \ww=\sqrt{\ww_{\mu\nu}\ww^{\mu\nu}/2 } \, ,\eqn{3.2}\fe 
by
 \be \ww_{\mu\nu}={1\over 2}\ww\, 
 \varepsilon_{\mu\nu\rho\sigma} {\calE}^{\rho\sigma}\, ,\eqn{3.3}\fe
where ${\cal E}^{\mu\nu}$ is the antisymmetric unit bivector 
(as normalised by ${\cal E}^{\mu\nu}{\cal E}_{\nu\mu}=2$)
tangential to the worldsheet -- which is uniquely defined modulo
the orientation convention involved in the choice of sign of the
spacetime measure tensor $\varepsilon_{\mu\nu\rho\sigma}$. There will be no
ambiguity of sign at all  in the specification of the
corresponding {\it fundamental tensor} of the worldsheet, 
namely the rank-2 tangential projection tensor that is given by
 \be \eta^\rho_{\ \sigma}={\cal E}^\rho{_\nu}{\cal E}^\nu{_\sigma} \, ,
 \eqn{3.4} \fe
nor in the complementary orthogonal projection tensor
\be \perp^{\!\rho}_{\, \sigma}=g^\rho_{\ \sigma}-\eta^\rho_{\ \sigma} 
\, ,\eqn{3.5}\fe
which will also be of rank-2. The latter will be definable directly
by
 \be \perp^{\!\rho}_{\, \sigma}=\ww^{-2}\ww^{\rho\nu}\ww_{\sigma\nu}\, 
 .\eqn{3.6}\fe

The satisfaction of the superfluid degeneracy requirement (\ref{3.1}) will
automatically hold as a consequence whenever $\ww_{\mu\nu}$ has a zero
eigenvalue characteristic vector such as exemplified by the current vector
$\nN{^\mu}$ when an equation of motion of the standard form (\ref{2.30a}) 
or its modification (\ref{2.30b}) is
satisfied. However, when the standard equation (\ref{2.30a}) is changed in the 
obvious way to
 \be \nN{^\mu}\ww_{\mu\nu}=\fNdrag{_\nu} \, ,\eqn{3.7}\fe
by the inclusion of a drag term $\fNdrag{_\nu}$, the  superfluidity
requirement (\ref{3.1}) will be violated unless the drag force has a very
special form, which will not be compatible with the usual kind of drag
proportional to the relative flow vector $\uC^\mu+ \uC{^\nu} \uN{_\nu}
\uN{^\mu}$. The obvious way to make sure that the force term on the right of
(\ref{3.7}) does not exclude the existence of a zero eigenvalue characteristic
for $\ww_{\mu\nu}$ is to postulate that it should be proportional to
$\ww_{\mu\nu}V^\mu$ for some vector $V^\nu$. The appropriate form for this
vector can  be deduced from the expression for the rate of entropy that
results from substitution of (\ref{3.7}) in (\ref{2.21}), which leads by
(\ref{2.27}) to
 \be \Theta\nabl_\mu s^\mu=\uC{^\nu}(\muN{_\nu} -\muC{_\nu})\nabl_\mu\nN{^\mu}
 +\uC{^\nu}\fNdrag{_\nu}-\uC{^\nu}\fE{_\nu} \, ,\eqn{3.8} \fe
in which, by the preceeding considerations, the drag contribution
$\uC{^\mu}\fNdrag{_\mu}$ will be proportional to $\uC{^\mu}\ww_{\mu\nu}V^\mu$.
Since any such internal contribution ought to be positive definite, in order
to satisfy the second law of thermodynamics, one is naturally lead to
postulate that the vector $V^\mu$ should itself be proportional to
$\ww_{\mu\nu}\uC{^\mu}$. In view of (\ref{3.6}) it follows that the
appropriate form for the drag force density on the superfluid will be given by
 \be \fNdrag{^\mu}=\drag\perp^{\!\mu}_{\, \nu}\uC{^\nu}\, ,\eqn{3.9}\fe 
for some positive resistivity coefficient $\drag$, that can be expected
to increase roughly in proportion to product of the vorticity magnitude
$\ww$ and the baryon density $\nC$ of the normal constituent. It is to
be remarked that a treatment involving a relativistic drag force
formula of this kind  has been previously developped in the context of
cosmic string theory by Vilenkin\cite{V91}\cite{CSM94}. In the specific
context of neutron star matter, a resistive drag formula interpretable
as the Newtonian limit kind described by (\ref{3.8}) has been obtained
on the basis of detailed microscopic analysis by Jones\cite{Jones90},
whose quantitative estimate for the coefficient $\drag$ in the lower
crust region will be discussed in Subsection \ref{6_2}, the main conclusion
being that it will be very small. This means that  in in the lower
crust region the zero drag limit, $\drag\rightarrow 0$ will provide
what for many purposes will be a very good aproximation, which will
described by the non-dissipative model governed by (\ref{2.30a}).  On
the other hand a more recent investigation\cite{SS95a} of conditions in
the high density core of the star indicates that the corresponding
value there will be very high.  This means that for this deep core
region so the opposite ``pinned'' limit, $\drag\rightarrow
\infty$, will provide a very good aproximation, which will be described
by the alternative non-dissipative model that is governed by
(\ref{2.30b}).

The analogous problem of the appropriate  form for the law governing the
superfluid creation rate $\nabl_\mu\nN^\mu$ is simpler because this creation
rate is just a scalar. It is evident from (\ref{3.8}) that the natural way to
ensure that this creation rate will be consistent with the second law of
thermodynamic is to postulate that it should be governed by a law of the form
 \be \nabl_\mu \nN^\mu=\chemr \uC^\nu(\muN{_\nu}-\muC{_\nu})\eqn{3.10}\fe
for some positive coefficient $\chemr$. Such a law is an obviously natural
generalisation of the kind of creation rate formula that is familiar
in chemical physics, but we are not aware of any microscopic analysis
providing an estimate of the appropriate value for $\chemr$. The situation is
complicated by the consideration that as far as the large scale mechanics of
the neutron star is concerned, the effective rate may depend not just on
microscopic processes, but also, when subduction is involved, on the rather
messy process whereby the crust is broken up before it ultimately dissolves.
In practice however it will suffice for many purposes, including the
application to be described below, to know that the effective value of the
chemical rate coefficient $\chemr$ is sufficiently high compared with the
relevant timescales of long term evolution for the local chemical equilibrium
condition (\ref{2.31b}) to be an adequate approximation as an alternative to
the more exact relation (\ref{3.10}), of which it represents the
non-dissipative limit as $\chemr\rightarrow\infty$. It is to be remarked that the
opposite limit $\chemr\rightarrow 0$ is also non-dissipative, providing a
non-transfusive treatment in which the superfluid constituent is separately
conserved according to (\ref{2.31a}) (as in the more familiar non-transfusive 
Landau type\cite{LK82}\cite{CK92}\cite{CL95a} of two fluid model) which
will be a good approximation for treating many high frequency processes
in neutron stars, but not so relevant for the long term evolution
processes to be considered here.

It is apparent from (\ref{2.21}) that the postulates (\ref{3.9}) and
(\ref{3.10}) can be combined in the single formula
 \be \fN{^\rho}= \drag\perp^{\!\rho}_{\, \nu}\uC{^\nu}+\chemr \muN{^\rho}
 \uC^\nu(\muN{_\nu}-\muC{_\nu})\, ,\eqn{3.11}\fe
for the force density $\fN{_\nu}$ acting on the superfluid 
constituent that is the primary subject of interest in the application to be
described below.  To complete the specification of the dynamical evolution
we would need an analogously explicit formula for the force density
acting on the normal constituent, which, by (\ref{2.23}) will have the form
 \be\fC{^\rho} =\fE^{\rho}-\fN^{\rho}\, ,\eqn{3.12}\fe
in which the external contribution $\fE{^\rho}$ still remains
to be specified. 

The simplest possibility is of course that in which the external force
contribution $\fE{^\rho}$ is absent, in which case the equations listed above
will be sufficient as they stand to determine the dynamical evlution. However
our purpose below is to consider cases in which the star is not effectively
isolated but subject to an external torque  that is ultimately
attributable to accretion or radiation reaction. Although it provided an
indispensible guide to the formulation of the explicit expression (\ref{3.11})
for the internal contribution $\fN{^\rho}$, the second law of
thermodynamics applies only to closed systems, so it does not provide any
information about the external contribution $\fE{^\rho}$: the final term
$\uC{^\nu}\fE{_\nu}$ in (\ref{3.8}) might have either sign depending on the
nature of the external force involved (broadly speaking one might expect it to
be positive in the case of accretion but negative for radiation reaction).

In order to cover a wide range of possible scenarios, the strategy
of the analysis below will be to refrain from adopting any specific
ansatz for the detailed distribution of the external force density
$\fE{^\rho}$, but to suppose that the evolution of the ``normal''
constituent is known in advance on the basis of other considerations,
so that if the value of $\fE{^\rho}$ were actually needed it could
just be read out from (\ref{3.12}). More specifically it will
be supposed in the following work that the motion of the ``normal''
constituent is approximately {\it rigid}. In view of its elastic solid
structure, this will obviously be a good approximation for the
crust, and in practice it will also be a very good approximation
in the deeper layers, which will be tightly coupled to the crust
by forces of various (particularly viscous and magnetic) kinds.
A more thorough treatment would need a detailed account of such
forces, whose presence will imply deviations from the perfectly
fluid description used here: this would be describable
in terms of adjustments
$\TE{^{\mu\nu}}\mapsto \TE{^{\mu\nu}}+\Delta T{^{\mu\nu}}$ and the
$\fC{^\rho} \mapsto  \fC{^\rho}+\Delta\fC{^\rho}$
with $\Delta\fC{^\rho}=\nabl_\nu(\Delta T{^{\nu\rho}})$,  but
would not directly affect the superfluid force density
$\fN{^\rho}$ in which the corresponding adjustment could be
neglected as a higher order correction. So long as we are not concerned with
the explicit form of the external force contribution, the effect of such
adjustments can be adequately taken into account using the non adjusted
formulae given above subject to the understanding that the quantity
$\TE{^{\mu\nu}}$ therein is to be interpreted as an {\it effective} energy
momentum tensor that differs from the {\it true} (but not so exactly known)
energy tensor $\TA{^{\mu\nu}}$ say by some adjustment expressible by
 \be \TE{^{\mu\nu}}=\TA{^{\mu\nu}}-\Delta T{^{\mu\nu}}\, ,\eqn{3.13}\fe
while similarly the quantity $\fE{^\rho}$ therein is to be interpreted as an
{\it effective} external force density that differs from the {\it true}
external force density, $\fA{^\rho}$ say, by a corresponding adjustment of the
form 
 \be \fE{^\rho}=\fA{^\rho}-\nabl_\mu\big(\Delta T{^{\mu\rho}}\big)
\, .\eqn{3.14}\fe
If the definition of the adjustment $\Delta T{^{\mu\nu}}$ is extended
to include all relevant radiation and/or accreting matter, then there
will remain no genuinely external force contribution, so the
first term $\fA{^{\rho}}$ in (\ref{3.14}) will simply disappear, 
i.e. the effective force density $\fE{^\rho}$ will be entirely
attributable to the adjustment term.

\section{Angular momentum distribution.}
\label{Sec3}

The main subject to which we wish to apply the foregoing formalism
on this occasion is the evolution of the angular momentum distribution
within a neutron star.

Before we can proceed it is to be remarked that in order for the useful
concepts of energy or angular momentum to be well defined in the strictest
sense, it is necessary that there should be a corresponding time stationnary
symmetry generator $\partial/\partial t=\kt^\mu\partial/\partial x^\mu$ say,
or an axisymmetry generator, $\partial/\partial\phi =\mh^\mu\partial/\partial
x^\mu$ say, whose action leaves the background spacetime structure invariant.
In a Newtonian treatment using a flat background a six parameter group of such
symmetries will always exist. However in a relativistic treatment using a
curved spacetime background, for which the relevant symmetry property is
locally expressible by the Killing condition that $\nabl_{(\mu}\kt_{\nu)}$ or
$\nabl_{(\mu}\mh_{\nu)}$ should vanish, no such solution will exist in the
generic case.

Fortunately, in the context of neutron stars, this problem of principle is
unimportant in practice. Although deviations from flat space geometry will
typically be rather large, several tens of percent, the relevant curved
spacetime geometry will nevertheless be able to be considered as time
independent to within a small fraction of a percent over the timescales during
which dynamical processes such as glitches occur, because the masses that are
set in motion by such process are very small compared with the relevant
Chandrasekhar limit: it will thus be possible to choose a time translation
generator $k^\mu$ such that  
\be \nabl_{\mu}\kt_{\nu}=\nabl_{[\mu}\kt_{\nu]}+\epsilon_{\mu\nu}\, ,
\eqn{4.0}\fe
in which the symmetric part $\epsilon_{\mu\nu} =\epsilon_{(\mu\nu)}$ is
sufficiently small to be neglected for most purposes: in formal language
$\epsilon_{\mu\nu} = {\cal O}\{L^{-1}\}$ where $L$ is a lengthscale that is
extremely large compared with the radius of the star.

In so far as axisymmetry is concerned the situation can be expected to be even
better: since the masses involved even in such conspicuously non axisymmetric
features  as a non aligned magnetic field will be extremely small, it will
usually be possible to treat the underlying spacetime geometry of a neutron
star as being effectively axisymmetric to a very good approximation in the
treatment of phenomena occuring not just on short and medium timescales, but
even over the very long timescales with which the present investigation is
chiefly concerned. In view of this, we shall proceed on the basis of the
supposition that there is a well defined axisymmetry generator $\mh^\mu$ say
that is characterised as an exact solution of the Killing equation
 \be \nabl_{(\mu}\mh_{\nu)}=0\, .\eqn{4.1}\fe

The presence of such a Killing vector allows us to define the angular
momentum, $\JE$ say, of the star at any instant by an integral of the
familiar form
 \be \JE=\int\calJE^\mu\, d\Si_\mu \ ,\eqn{4.2}\fe where the local
angular momentum current is defined by
 \be \calJE^\mu=\mh^\nu\TE{^\mu}_\nu \, ,\eqn{4.3}\fe and the integral
is taken over a spacelike hypersurface $\Si$ characterising the instant
under consideration, using the convention that the normal surface
element covector $d\Si_\mu$ in the integrand is directed towards the
{\it past} (in order to avoid the introduction of the minus sign that
would be needed for the more usual future directe orientation
convention).  Although more general kinds of spacelike hypersurface
might be envisaged, it will be taken for granted throughout the
discussion that follows that in order to be admissible for the purpose
of this definition the hypersurface must itself be invariant under the
axisymmetry action, which means that with respect to the Killing vector
the normal element $d\Si_\mu$ must satisfy the tangentiality condition
 \be  \mh^\mu\,d\Si_\mu=0\, .\eqn{4.4}\fe

When the hypersurface used to evaluate the angular momentum is subject to the
action of a time translation generator $\partial/\partial
t=\kt^\mu\partial/\partial x^\mu$ say, the corresponding rate of variation,
for which we shall use the usual abbreviation
 \be \dot{\JE}={d\JE\over dt}\, , \fe
will be given by the identity
 \be \dot{\JE}=\int\big( {\Libra}_{\kt}\calJE{^\mu} +\calJE{^\mu}
 \nabl_\nu \kt^\nu\big)\, d\Si_\mu \, ,\eqn{4.5}\fe
in which the Lie derivative is just the commutator
 \be  {\Libra}_{\kt}\calJE{^\mu} =\kt^\nu\nabl_\nu\calJE{^\mu}
-\calJE{^\nu}\nabl_\nu \kt^\mu\, ,\eqn{4.6}\fe
and the divergence contribution $\nabl_\nu \kt^\nu$ would vanish if the vector
$\kt^\mu$ were taken to be an exact solution of the Killing equation, 
as would be possible if the background were exactly stationary, an assumption
which will not be needed for our present purpose. 

Independently of whether or not $\kt^\mu$ is actually a Killing vector, the 
standard time variation formula (\ref{4.5}) can be identically rewritten in
the convenient form
 \be \dot{\JE}=\int \big(\kt^\mu \nabl_\nu\calJE{^\nu}\, 
 +2\nabl_\nu(\calJE{^{[\mu}} \kt{^{\nu]}})\big)\, d\Si_\mu\, ,\eqn{4.7}\fe
in which the last term is a divergence. This means that its contribution can
be converted, by the Green theorem, into a two-dimensional integral over the
boundary ${\cal S}=\partial\Si$ of the spacelike hypersurface, giving the
identity 
 \be  \dot{\JE}=\int \kt^\mu \nabl_\nu\calJE{^\nu} \, d\Si_\mu
 +\oint \calJE{^{[\mu}}\kt{^{\nu]}} \,d{\cal S}_{\mu\nu}\, ,\eqn{4.8}\fe 
where $d{\cal S}_{\mu\nu}$ is the (timelike) normal 2-surface element of the
(spacelike) boundary.

In the physical application with which we are concerned here, the boundary in
(\ref{4.8}) can be taken outside the surface of the neutron star so that its
contribution drops out, and it follows from the Killing equation (\ref{4.1})
and the definition (\ref{2.24}) of the external force density $\fE_{_\mu}$ that
the divergence of the angular momentum current (\ref{4.3}) will be given by
\be \nabl_\mu\calJE{^\mu}=\mh^\nu\fE{_\nu}\, .\eqn{4.9}\fe
The rate of change of the total angular momentum is thus finally obtained
in the form
 \be \dot{\JE}={\mit\Gamma}\, ,\eqn{4.10}\fe
where ${\mit\Gamma}$ is the total external torque as given by
 \be {\mit\Gamma}=\int\mh^\nu \fE{_\nu}\kt^\mu\, d\Si_\mu\, .\eqn{4.11}\fe

Subject to the restriction (\ref{4.4}), it can be seen from the form of the
expression (\ref{2.18}) for the total energy momentum tensor that -- despite
the fact that we are allowing for the possibi lity that the superfluid
constituent may interact strongly with the crust and the rest of the
``normal'' material that is dragged along with it -- there will nevertheless
be an unambiguous decomposition of the total angular momentum (\ref{4.2}) as a
sum of the form
 \be \JE=\JC+\JN\, ,\eqn{4.13}\fe
in which the ``normal'' contribution $\JC$ 
(due mainly to the crust) is given by
 \be \JC=\int\calJC^\mu\, d\Si_\mu \, ,\eqn{4.14}\fe
with 
 \be \calJC^\mu=\mh^\nu\big(s\Theta\uC{^\mu}\uC{_\nu}
 +\nC{^\mu}\muC{_\nu}\big) \, ,\eqn{4.15}\fe
while the superfluid contribution $\JN$ is given by
\be \JN=\int\calJN^\mu\, d\Si_\mu \, ,\eqn{4.16}\fe
with 
 \be \calJN^\mu=\aN \nN^\mu \, ,\eqn{4.17}\fe
where $\aN$ is the angular momentum per superfluid neutron as
defined simply by
\be \aN=\mh^\nu\muN{_\nu} \, .\eqn{4.18}\fe

It is to be remarked that the local angular momentum current will have
the form
 \be \calJE^\mu=\calJC^\mu+\calJN^\mu+\calJh^\mu \, ,\eqn{4.19}\fe in
which there will be an extra term
 \be \calJh^\mu=\Psi\mh^\mu \,  ,\eqn{4.21}\fe that cannot be
unambiguously decomposed into collectively comoving and superfluid
parts except in the separable limit for which the Lagrangian introduced
in (\ref{2.6}) can itself be decomposed in the form
$\Lamb=-\rhoC-\rhoN$, which, as already mentionned, may be an
acceptable approximation at moderate densities, but is unlikely to be
accurate in the deeper regions. It can be seen however that -- provided
(\ref{4.4}) is respected -- the mongrel term (\ref{4.21}) will not
contribute to the integrated total.

It can be seen from the definition (\ref{2.21}) of the force density
$\fN{_\mu}$ acting on the superfluid constituent that the divergence of its
angular momentum contribution will be given identically by
 \be \nabl_\mu\calJN{^\mu}=\mh^\nu\fN{_\nu}+ \nN{^\nu}{\Libra}_{\mh}\muN{_\nu}
 \, ,\eqn{4.22}\fe
in which the Lie derivative 
 \be {\Libra}_{\mh}\muN{_\nu}=\mh^\rho\nabl_\rho\muN{_\nu}+\muN_\rho
 \nabl_\nu\mh^\rho \eqn{4.23}\fe
will vanish provided the stellar configuration itself (and not just the
background spacetime as has been assumed so far) is invariant under the
axisymmetry action, in which case -- by the same reasonning used to obtain
(\ref{4.10}) -- the rate of change of $\JN$ will be given simply by
 \be \dot{\JN}=\int\mh^\nu \fN{_\nu}\kt^\mu\, d\Si_\mu\, .
 \eqn{4.24}\fe

It can similarly be seen from the definition (\ref{2.20}) of the force density
$\fC{_\mu}$ acting on the ``normal'' constituent that the divergence of its
angular momentum contribution will be given identically by
 \be \nabl_\mu\calJC{^\mu}=\mh^\nu\fC{_\nu}+ \nC{^\nu}{\Libra}_{\mh}\muC{_\nu}
 +s^\nu{\Libra}_{\mh}(\Theta \uC_\nu) \, .\eqn{4.25}\fe
Here again the Lie derivatives involved will vanish provided the stellar
configuration shares the axisymmetry property of the background,
in which case the rate of change of $\JC$ will be given by
the obvious analogue of (\ref{4.24}), namely
\be \dot{\JC}=\int\mh^\nu \fC{_\nu}\kt^\mu\, d\Si_\mu\, . \eqn{4.26}\fe

\section{Evolution equations.}
\label{Sec4}

For the purpose of keeping account of the evolution of the angular momentum
distribution, a particularly handy quantity to work with is the angular
momentum per superfluid neutron, $\aN$, that was introduced in (\ref{4.18}),
since the axisymmetry requirement to the effect that the Lie derivative
(\ref{4.23}) should vanish, is expressible as a formula giving the gradient of
$\aN$ in terms of the vorticity in the form 
 \be \nabl_\mu\aN=\ww_{\mu\nu}\mh^\nu \, .\eqn{5.1}\fe

This relation is convenient for processing the basic superfluid
equation of motion, which will be given, according to (\ref{3.7})
and (\ref{3.9}), by
\be \nN{^\mu}\ww_{\mu\nu}= \drag\perp_{\nu\sigma}\uC{^\sigma}
\, ,\eqn{5.2}\fe
which can be rewritten using (\ref{3.6}) and the orthogonality property
$\ww_{\mu\rho}\perp^{\!\rho}_{\, \nu}=\ww_{\mu\nu}$ in the equivalent 
alternative form
\be \ww^2\nN{^\nu} \perp^{\!\rho}_{\, \nu}=\drag\ww^{\rho\nu}\uC_\nu
\, .\eqn{5.3}\fe
Contracting these with the axial Killing vector $\mh^\mu$ one obtains
a pair of dynamical equation that take the forms
\be\nN^\nu\nabl_\nu\aN=\drag\ph{^\nu}\uC_\nu\, ,\eqn{5.4}\fe
and
\be \ww^2\nN{_\nu}\ph{^\nu} =-\drag\uC{^\nu}\nabl_\nu\aN
\, ,\eqn{5.5}\fe
using the abbreviation 
\be \ph{^\mu}= \perp^{\!\mu}_{\, \nu}\mh^\nu\, ,\eqn{5.6}\fe
for the vortex sheet orthogonal projection of the Killing vector.

As already mentionned, it will be a very good approximation for our
present purpose to suppose that the motion of the ``normal''
constituent is very nearly rigid, so that its unit flow vector will be
expressible in the form
 \be \uC{^\mu}=\gamC(\kt{^\mu}+\OmC\mh{^\mu})\, , \eqn{5.8}\fe where
$\gamC$ is a Lorentz type factor allowing for gravitational and Doppler
redshifts, and where $\OmC$ is  a {\it uniform} angular velocity
(representing what is actually observed in pulsars) and $\kt^\mu$ will
be an approximate solution of the Killing equation, as characterised by
(\ref{4.0}). With respect to the exact Killing vector $\mh^\mu$ and the
approximate Killing vector $\kt^\mu$ that is characterised by this
approximate rigidity condition, the unit flow velocity $\uN{^\mu}$ of
the superfluid will be expressible in the analogous form
 \be \uN{^\mu} =\gamN(\kt{^\mu}+\OmN\mh{^\mu}+\vN^\mu)\, , \eqn{5.9}\fe
in which the extra term $\vN^\mu$ allows for the possiblity of a small
non circular (convective) motion which can be expected to be very small
under the conditions of interest here, but in which the most important
difference from (\ref{5.8}) is that the superfluid angular velocity
$\OmN$ defined by (\ref{5.9}) is not supposed to be even approximately
uniform.  (We have been able to avoid the need to include an
analogous small convection velocity term in (\ref{5.9}) by taking
advantage of the fact that we are not supposing that $k^\mu$ is an
exact Killing vector, which means that there is some gauge freedom in
its specification:  the understanding here is that this gauge freedom
has been used to absorb the small convection correction that would
otherwise have been needed in (\ref{5.8}), so that this equation is
interpretable not just as a statement about the approximate rigidity of
the ``normal'' constituent but also as a gauge fixing condition for
$k^\mu$.)

In terms of the uniform angular velocity $\OmC$ and the variable angular
velocity $\OmN$ the information contained in (\ref{5.4}) and (\ref{5.5}) is
expressible, using the abbreviations
\be \dot{\aN}=\kt^\mu\nabl_\mu\aN\, ,\hskip 1 cm
 \ph^2=\ph{^\mu}\ph{_\mu}\, ,\eqn{5.10}\fe
 by  the pair of equations
\be -\ph^{-2}\kt_{_\mu}\ph{^\mu}={\OmN+\drac^2\OmC+\Omminus
\over 1+\drac^2}\, ,\eqn{5.11}\fe
\be \ww^{-1}\ph^{-2}\dot{\aN}={\drac\big(\OmC-\OmN-\Omplus\big)
\over 1+\drac^2} \, ,\eqn{5.12}\fe
in which $\drac$ is the dimensionless drag coefficient given by
\be \drac={\drag\gamC\over
\ww\nN\gamN} \, ,\eqn{5.13}\fe
where the convection  contributions  $\Omplus$ and
$\Omminus$, expressed dimensionnally as angular velocity corrections,
 will be sufficiently small to be neglected for most purposes, their
explicit values being given by 
\be \Ompm=\ph^{-2}\vN{^\nu}\big(\ph{_\nu}\pm\drac^\mp\ww^{-1}\nabl_\nu\aN\big) \,
.\eqn{5.14}\fe

The term on the left of the first expression (\ref{5.11}) can be interpreted
as the angular velocity of the vortex array: one sees that it represents a
weighted mean of the angular velocities of the normal and superfluid
constituents. This is not surprising because each fluid acts so as to minimise
the relative velocity between the vortices and itself. There are two extreme
cases: in the limit $\drac\rightarrow 0$, when the resistive drag is
very small, the vortex array ``feels'' only the superfluid and co-rotates with
it; in the opposite limit $\drac\rightarrow \infty$, when the drag coefficient is
very high, the vortex array co-rotates with the rigidly rotating ``normal''
constituent. 

In view of (\ref{5.1}) the term on the left of the second expression
(\ref{5.12}) can be interpreted in a similar way as representing the
non-circular ``convective'' component of the velocity of the vortices
in a direction orthogonal both to their direction of alignment and to
the axisymmetry generator $\mh^\mu$, but the reason why it is of
particular interest for our present purpose is that it provides the
value of the quantity $\kt^\nu\nabl_\nu\aN$ that measures the rate at
which $\aN$ (the angular momentum per superfluid neutron) changes with
time. As one would have expected, it is roughly proportional to the
difference $\OmC-\OmN$ with a coefficient that is large only for
intermediate values of $\kap$. It is not surprising that the
coefficient is small when $\drac$ is small since in this case the
superfluid hardly ``feels'' the ``normal part'' at all. The paradox
that the coefficient is {\it also small} for very high values of
$\drac$, i.e. when the drag force is strong, can be explained as due to
the fact that in this case the friction prevents the development of the
transverse ``convective'' motion of the vortices: in such circumstances
the Joukowski force due to the Magnus effect will remains orthogonal to
the direction of rotation, which renders it ineffective for reducing
the difference of the rotation speeds.

The quantity given by (\ref{5.2}) does not quite constitute the entire
superfluid force density $\fN{_\mu}$ that is required for evaluating   the
integral in (\ref{4.24}), since it does not include the contribution in the
complete expression (\ref{3.11}) allowing for the possible creation or
destruction of the superfluid material. According to (\ref{3.11}) the required
torque density will be given by
 \be \mh{^\rho}\fN{_\rho}= \drag\ph{^\rho}\uC{_\rho}+
\chemr\mu_\rho \mh{^\rho}
 \uC^\nu(\muN{_\nu}-\muC{_\nu})\, .\eqn{5.16}\fe
Evaluating this for the configuration described by (\ref{5.8}) and
(\ref{5.9}), using the results that have just been obtained, gives the
superfluid torque density in the more explicit form
 \be \mh{^\rho}\fN{_\rho} ={\drac\gamN\nN\ww\ph^2\over 1+\drac^2}
\big(\OmC-\OmN-\Omminus\big)
+\aN\chemr\uC{^\nu}(\muN{_\nu}-\muC{_\nu}) \, ,\eqn{5.17}\fe
in which the chemical adjustment term proportional to $\chemr$,
and the term $\delp$ allowing for convection can be expected to be very
small compared with the dominant contribution which is proportional
to the local angular velocity difference $\OmC-\OmN$. 

The system of equations that is thus obtained is a relative
generalisation of a system of the kind that is already familiar in the
Newtonian equation \cite{Jones90}.  In the slowly rotating limit, the
angular momentum contributions that we have been considering will be
able to be treated just as homogeneous linear combinations of the
relevant angular velocity variables, while the convection terms
proportional to $\vN{^\mu}$ and the chemical adjustment terms
proportional to $\chemr$ will be able to be neglected altogether. It
can be seen that in this limit the evolution of the relevant angular
velocity variables, under the influence of a weak arbitrarily time
dependent external torque ${\mit\Gamma}$, will be completely determined
by the equations that have just been obtained.  The way this works is
particularly transparent in the separable case (meaning the case in
which the entrainment effect mentionned above is neglected, which is 
strictly true only in the crust where there are no superconducting 
protons)
 for which the Lagrangian is decomposible in the  form
$\Lamb=-\rhoC-\rhoN$ that was mentionned above, since in this case the
angular momentum per superfluid neutron, $\aN$, at any position will
simply be proportional to the local value of the superfluid angular
velocity $\OmN$ there, while the angular momentum $\JC$ of the entire
rigidly rotating ``normal'' constituent will be proportional just to
the single uniform angular velocity $\OmC$ that is directly observable
from outside. The time evolution of $\aN$ is given by (\ref{5.12}),
while the time evolution of $\JC$ is obtainable by substituting
(\ref{5.17}) in the equation that is obtainable via (\ref{4.10}) and
(\ref{4.13}) from (\ref{4.26}) in the form
 \be \dot{\JC}={\mit\Gamma}-\int\mh^\nu \fN{_\nu}\kt^\mu\, d\Si_\mu\, .
 \eqn{5.18}\fe
The same system of linear equations -- namely (\ref{5.12}) and the
substitution of (\ref{5.17}) in (\ref{5.18}) -- will still be sufficient to
determine the evolution of the angular velocities for a slowly rotating system
in the non-separable case, the only difference being that the matrix linearly
relating the angular velocity variables to their time derivatives will have
more numerous off-diagonal components: in the generic case $\aN$ will no
longer be proportional just to $\OmN$ but to a linear combination of $\OmN$
with $\OmC$, while $\JC$ will no longer be proportional just to $\OmC$ but to
a linear combination of $\OmC$ with some appropriately weighted linear average
of the distribution of $\OmN$ over the star. It is to be noticed 
however that, because of the expected small density of superconducting
protons with respect to superfluid neutrons in the core (a few percents),
 the terms due to the non-separability of the Lagrangian will remain
small.

\section{Estimation of the relevant orders of magnitude.}
\label{Sec6}

Before considering the effects that were left out, in order to determine the
circumstances under which their neglect is justified, we  shall first
estimate the magnitude of the two main kinds of force, namely the Joukowski
force and the friction drag force on the vortices,  that  effectively governs
the motion of the material of the neutron star in the treatment that has just
been described.

\subsection{ The Joukowski-Magnus lift force}
\label{6_1}

As soon as there is a relative motion between a thin elongated structure such
as an aerofoil or a superfluid vortex and the ambient fluid, the Magnus effect
produces a non-dissipative ``lift'' force that is orthogonal to the relative
motion. According to the
well known formula due to Joukowski, the magnitude of the lift force per unit
length, in any irrotational fluid or superfluid background, will simply be
proportional to the relevant momentum circulation integral. (Joukowski's
theorem was originally developped for application to aerofoils in the
terrestrial atmosphere, but it is sufficiently robust to remain valid even in
a highly relativistic context). The Joukowski formula is particularly
convenient for application in the context of superfluids, in which the
relevant momentum circulation integral, $\oint\muN{_\nu}\, dx^\nu$, will be given
in advance just by Planck's constant or a simple fractional multiple thereof.
(In the context of aerofoil theory the estimation of the value of the
circulation -- not to mention its control, which is the secret of success in
flying -- is not quite so easy).

In ordinary liquid Helium, the momentum circulation integral will just be the
Planck's constant $h=2\pi\hbar$ if the current is measured in terms of entire
Helium atoms, of which each contains four baryons, but if the current is
measured in baryon units, whose average momentum will be a quarter of that of
a whole atom, the corresponding circulation integral will be just $h/4$. In
the present application the analogue of the Helium atom is a Cooper type {\it
pair} of neutrons, which means that as we have chosen to measure the current
in baryon units the relevant momentum circulation constant will be given by
$\oint\muN{_\nu}\, dx^\nu=h/2=\pi\hbar$. For a circle of radius $r$,
orthogonal to  a small bunch of included vortex lines, the momentum
circulation can be evaluated as $\pi r^2\ww$, which means that the vorticity
scalar $\ww$ is interpretable as representing the momentum circulation per
unit area. It follows that for a superfluid constituted by neutron pairs with
the current measures in baryon units the number density of vortex lines per
unit area of an orthogonal section will be given by $\ww/\pi\hbar$. 

According to the left hand side of formula (\ref{3.7}), which represents 
the relativistic version of the Joukowski-Magnus force, the magnitude
$\fMagnus$ of the ``lift'' force per unit volume due to a relative flow
velocity $v_{_{\rm nv}}$ orthogonal to the vortices of the superfluid neutrons
will be given in terms of their number density $\nN$ by $\fMagnus=\nN\ww\,
v_{_{\rm nv}}$. The corresponding magnitude $\FMagnus$ of the ``lift'' force
per unit length on an individual vortex will therefore be given simply by
 \be \FMagnus= \pi\hbar \nN \, v_{_{\rm nv}} \, ,\eqn{6.12}\fe
which, in the non relativistic limit, can be seen to be in satisfactorily
perfect agreement with what is given by the classical Joukowski ``lift''
force formula. (Note that the factor $\pi\hbar$ would have to be raplaced by
$h=2\pi\hbar$ if one wanted to interpret $\nN$ as the number density of neutron
pairs, not just the number density of individual neutrons as is done here.)

\subsection{The resistive drag force}
\label{6_2}

For a relative flow velocity $v_{_{\rm cv}}$ of the corotating ``normal''
constituent orthogonally to the vortices the magnitude $\fdrag$ of the
resistive drag force per unit volume will be given according to our formula
(\ref{3.9}) by $\fNdrag=\drag v_{_{\rm cv}}$. Dividing this by the vortex
number density per unit area, $\ww/\pi\hbar$, as before, one sees that the
corresponding expression for the force per unit length $\Fdrag$ on an
individual vortex line will be given by
 \be \Fdrag= {\pi\hbar\drag\over\ww} v_{_{\rm cv}}
 \simeq\pi\hbar\nN\drac\, v_{_{\rm cv}}\, ,\eqn{6.14}\fe
where $\drac$ is the dimensionless resistivity coefficient introduced in
(\ref{5.13}).
             
In the context with which we are concerned, the force by which this resistive
drag is balanced will be mainly provided by the Joukowski lift due to the
Magnus effect. Comparing (\ref{6.12}) and (\ref{6.14}) it can be seen that
under such conditions, the magnitude ratio of the mutually orthogonal relative
velocities of the corotating ``normal'' matter and the superfluid with respect
to the vortices will be given by
 \be {v_{_{\rm nv}}\over   v_{_{\rm cv}} } ={\drag\over\ww\nN}
 \simeq{\drac}\, ,\eqn{6.16}\fe
from which it can be seen that $\drac$ is interpretable as the relativistic
generalisation of the drag to lift ratio that is the tangent of what is known
in classical aviation theory as the ``gliding angle''.

Using the rough order of magnitude estimates $\ww\approx 2\mN\OmN$, for
the vorticity, and $\aN\approx\mN\OmN\varpi^2$ for the angular momentum per
superfluid neutron, where $\varpi$ is a cylindrical radial coefficient, in
terms of which we shall also have $\ph\approx\varpi$, it can be seen that the
equation (\ref{5.12}) provides a rough estimate for the rate of change of the
local superfluid angular velocity in the form
 \be {\dot\OmN\over\OmN}\approx {2\drac\big(\OmC-\OmN\big)\
 \over 1+\drac^2}\, ,\eqn{6.17}\fe
which is interpretable as meaning that the superfluid response timescale,
$\tau$,  to a change of the crust angular velocity, i.e. the characteristic
lifetime for survival of a local angular velocity deviation $\OmN-\OmC$
against resistive damping, will be roughly given by
 \be \tau\approx {_1\over^2}(\drac+\drac^{-1})\vert\OmC-\OmN\vert^{-1}\,
 .\eqn{6.18}\fe
The application of this formula requires knowledge of the local value
of just a single parameter, namely the drag ratio $\drac$. 

It can be seen that the timescale $\tau$ is shortest when $\drac$ is of the
order of unity, in which case the drag force will be of the same magnitude as
the Joukowski force given by (\ref{6.12}). However it is likely that $\tau$
will greatly exceed the lower limit $\vert\OmC-\OmN\vert^{-1}$, not only in
the lower crust where one  expects\cite{Jones90} $\drac$ to be small, and also
in the inner core where recent investigations\cite{SS95a} suggest that $\drac$
is likely to be very large compared with unity.

In the lower crust region, to which the most detailed studies have been
devoted, an estimate of the relevant drag coefficient has been
provided by the work of Jones\cite{Jones90}, who predicts that it should be
proportional to the inverse fifth power of the relevant pairing correlation
length $\xiN$ say, which roughly characterises the radius of the vortex cores.
The value of this quantity is a sensitive function of density, with a
dependence that is still subject to a considerable degree of theoretical
uncertainty. For the lower crust region that is most important for the kind of
application under consideration here, typical estimates\cite{PA91} are in the
range $\xiN\gta 10^3\kFermi^{-1}$ where $\kFermi$ is the Fermi wave number of
the superfluid neutrons, which is related to their number density by
$\kFermi^3=3\pi^2\nN$. Another way of expressing this is to say that the
corresponding pairing energy gap, $\DelP\simeq \hbar^2\kFermi/\mN\xiN$, is in
the range $\DelP\lta 10^{-3}\EFermi$, where $\mN$ is the neutron mass and
$\EFermi\simeq \hbar^2\kFermi^2/2\mN$ is the Fermi energy of the neutron
superfluid.

The Jones formula is expressible as the statement that, in the lower crust 
region, the drag ratio $\drac$ will be given by
 \be \drac\simeq{3\aC\EPin^2\xiN^{-3}\over 
 32\pi^{3/2} \hbar\NC\mN\nN\csound^3}\, ,\eqn{6.20}\fe
where $\aC$ is the lattice spacing lengthscale,  characterising the mean
separation between the ionic crust nuclei, in which the non superfluid baryons
are concentrated, and $\NC\approx\nC\aC^3$ is the number of baryons (neutrons
and protons) per nucleus, while $\csound$ is the phonon speed in the neutron
superfluid and finally $\EPin$ is the ``pinning'' energy by which any crust
nucleus located within a vortex core is bound. According to standard results
developed by Alpar, Anderson, Pines and Shaham \cite{AAPS84a}\cite{AAPS84b},
and summarised by Ruderman\cite{R91}, the latter will be given by 
 \be \EPin\simeq{\hbar^2\aNucl^2\nN\over\pi\mN\xiN}\, ,\eqn{6.22}\fe 
where $\aNucl$ is the radius of the crust nucleus, which will
be given roughly by $\aNucl\approx 10\,\hbar \NC^{1/3}/\mN$.   Combining
(\ref{6.20}) and (\ref{6.22}), the Jones formula is obtained in the form
 \be\drac\simeq{3\hbar^3\aC\aNucl^4\nN\xiN^{-5}
 \over 32\pi^{7/2}\NC\mN^3\csound^3}\, .\eqn{6.24}\fe
It is not easy to draw precise quantitative conclusions from this formula
because of the high degree of theoretical uncertainty about the density
sensitive correlation length $\xiN$ which comes in at an inverse fifth power,
but since the other lengthscales involved will presumably be smaller than or
-- in the case of the internuclear spacing $\aC$ -- at most comparable with
$\xiN$, and since the phonon speed will be quite high, it seems clear that the
outcome will always be small, and that it will typically be very small,
$\drac\ll 1$. 

The state of affairs in the inner core is very different. According to a
detailed investigation that has recently been carried out\cite{SS95a}, the
main resistive drag force in the protonically superconducting superfluid below
the crust is due to the non-separability property, and the consequent
entrainment, which results in the trapping of large numbers of protonic
vortices by each neutron superfluid vortex: the estimated value of the
resistivity due to electron scattering by these very tiny magnetised flux
tubes is expressible by the formula
 \be \drag={\pi\hbar\, \kFerme^2\,\penetr^{3-|\entra|}\,\xiP^{|\entra|} \over
 \sqrt 3\, 2^6\, |\entra|} \, ,\eqn{6.30}\fe 
in which  $\kFerme$ is the Fermi wave number of the degenerate electrons,
$\xiP$ is the pairing coherence length of the superconducting protons,
$\penetr$ is the magnetic penetration lengthgscale, which will be given very
roughly by $\penetr^2\approx \mN c^2/4\pi e^2\nC$ where $e$ is the proton
charge, and finally the index $|\entra|$ is given by $\entra\simeq\mP/\DelmP$
where $\DelmP$ is the deviation (due to the entrainment) of the effective mass
of the proton from its usual value $\mP$, for which the numerical value is
thought\cite{Sj76} to be somewhere in the range $-5\lta \entra \lta -2$. As in
the previous example, the uncertainty in the relevant correlation lenghtscale,
in this case $\xiP$, which comes in at a high and itself uncertain power,
makes it hard to draw precise quantititative conclusions from (\ref{6.30}).
Nevertheless, since it involves only quantities at microscopic nuclear
physical scales, and since division not just by the (nuclear order) number
density but also by the macroscopic vorticity $\ww$ (which will be very small
by nuclear standards) is required to obtain corresponding dimensionless drag
ratio, $\drac$ it is clear that the latter will always turn out to be
extremely large, $\drac\gg 1$.

\section{Range of validity of the analysis.}
\label{Sec7}

Before concluding we shall try to form an idea of the range of conditions
under which the treatment that has just been developped should be valid,
at least as a first order approximation. We shall do this by estimating
the relative orders of magnitude of potentially important effects that
have not been taken into account as compared with those that are
treated as dominant in the analysis above.

\subsection {The tension force}
\label{7_1}

Among the potentially significant effects that were {\it not} taken
into account in the treatment provided here, that first to which we
shall address our attention is the deviation of the stress momentum
energy contribution of the superfluid from a perfect fluid form due to
the effective tension, $T$ say, arising from the string-like nature of
the vortices.  An elegant relativistic formalism for the description of
this effect has recently been made available \cite{CL95c} but it turns
out not to be needed for the analysis of the very large scale long term
evolution that is considered here. Although the effect of the vortex
tension may in certain circumstances become important at a local level,
it tends to be negligible for large scale phenomena because the
associated force per unit length depends on the bending of the vortex
lines, and is proportional to the magnitude $\vert K\vert$ of the
curvature vector $K_\rho$ of the string worldsheet.  For the large
scale effects with which we are concerned here the relevant curvature
radius $\Rcurv=\vert K\vert^{-1} $ can typically be expected to be of
the order of the thickness of the layers involved, and thus comparable
with a not insignificant fraction of the radius of the star as a
whole.  The tension force per unit length $\Ftension$ on an individual
vortex has been shown\cite{CL95c} to be expressible in the form
 \be {\Ftension}_\rho= T K_\rho-\perp^\sigma_\rho\nabla_\rho T\,
 .\eqn{7.1} \fe
The geometric curvature vector $K_\rho$  of the vortex worldsheet is
expressible in terms of its fundamental tensor by
 \be K_\rho=\eta^\nu_{\ \sigma}\nabl_\nu\eta^\sigma_{\ \rho}= {\cal
 E}^\nu_{\ \sigma}\nabl_\nu{\cal E}^\sigma_{\ \rho}\, . \eqn{7.2}\fe
The vortex tension $T$ will be given\cite{CL95b} by the formula
 \be T={\pi\hbar^2\over 2}{\nN\over \muN}\, {\rm ln}\Big\{
 {\ww_{_\odot}\over \ww}\Big\} \, ,\eqn{7.3} \fe 
in which $\ww_{_\odot}$ is a fixed parameter interpretable as the maximum 
order of magnitude that would be attained by the vorticity magnitude $\ww$ 
in the limit when the vortex cores are almost in contact, so that in terms
of the correlation length $\xiN$ introduced in Subsection \ref{6_2},
which provides an estimate of the vortex core radius, it will be given
by $\ww_{_\odot}\approx \hbar\xiN^{-2}$.  This means that the
ratio $\ww_{_\odot}/ \ww$ can be evaluated as the square of the ratio
of the mean intervortex separation distance to the vortex core radius
$\approx\xiN$, so for the  rotation rates
typical of neutron stars the logarithmic factor will be fairly large,
${\rm ln}\big\{ \ww_{_\odot}/ \ww\big\}\approx 40$.
Since this factor has only a very weak dependence on $\ww$, the
gradient term in (\ref{7.1}) will be relatively negligible.  For a
rough order of magnitude estimation the effective mass $\muN$ per
superfluid neutron can be taken to  be given by its Newtonian limit
value $\muN \simeq \mN$, so  the amplitude of the tension force is
found to be given by
 \be \Ftension\approx {20\pi \hbar^2\nN\over
 \mN \Rcurv} \, . \eqn{7.4}\fe 
The ratio of this tension force to the Magnus lift force $\FMagnus$
given by the Joukowski formula (\ref{6.12}) will therefore be given by
\be {\Ftension\over\FMagnus}\approx{20\hbar\over
 \mN v_{_{\rm nv}}\Rcurv} \, ,\eqn{7.5}\fe
where $v_{_{\rm nv}}$ is the flow velocity of the superfluid neutrons
relative to the vortices and hence, as we have seen, relative to the
corotating ``normal'' matter in the core. For typical differential
angular velocities of a few rotations per second the relative velocity
will be of the order of $10^{-4}$ in units such that the speed of light
is unity. Assuming that the relevant bending radius  $\Rcurv$
represents  a significant fraction  of the stellar radius (of the
order of $10^{-19}$ times larger than the neutron Compton radius) we
see that in the stellar core the ratio (\ref{7.5}) will be given by
${\Ftension/\FMagnus}\approx 10^{-15}$, which means that the tension
force will indeed be entirely negligible as we have been assuming.  

In contrast with the case of the core, the situation in the crust of
the star is rather more delicate, since we have seen that $v_{_{\rm
nv}}$ will be much smaller than the relevant differential rotation
velocity there, which will be given by $v_{_{\rm cv}}$, i.e. by the
relative flow speed of the crust material relative to the vortices. The
latter determines the drag force $\Fdrag$ according to the formula
(\ref{6.14}) which gives
 \be {\Ftension\over\Fdrag}\approx{20\hbar\over \mN \drac v_{_{\rm
 cv}} \Rcurv} \, .\eqn{7.6}\fe \ 
For typical differential angular velocities of a few rotations per
second in the lower crust we obtain a numerical estimate of the form
${\Ftension/\Fdrag}\approx 10^{-15}\drac^{-1}$.  In order to obtain the
strong inequality, $\Ftension\ll\Fdrag$, that we want in order to
justify our neglect of the tension force in the crust as well as in the
core, all we need is to be sure that although it is very small compared
with unity, the drag ratio $\drac$ will nevertheless satisfy
$\drac\gg 10^{-15}$. This does however seem a fairly safe conclusion to
draw from the Jones formula (\ref{6.20}), despite the considerable
degree of uncertainty in the evaluation of the relevant microscopic
parameters.

\subsection{Vortex pinning}

The general formalism set up in Section\ref{Sec2} is perfectly capable of
treating vortex pinning, for which it suffices to use the non-dissipative
equation of motion (\ref{2.30b}) that represents the limit
$\drac\rightarrow\infty$ of the generic equation of motion (\ref{5.2})
provided by  (\ref{3.7}) and (\ref{3.9}).  As remarked above the ``pinned
limit'' equation of motion (\ref{2.30b}) will provide a very good approximation
in the core region where we have seen that  the drag ratio $\drac$ can be
expected to be extremely high. However although its macroscopic effect is
similar, from a microscopic point of view the high drag effect that is
predicted in the core is very different from the stricter kind of pinning
whose ocurrence in the crust was originally proposed by Anderson and
Itoh\cite{AI75} as a mechanism that might explain the very large glitches
observed in the Vela pulsar. Pinning in this strict sense is presumed to occur
as a consequence of the attraction characterised by the binding energy $\EPin$
given by (\ref{2.26}) that occurs between a superfluid vortex core and
a ionic nucleus in the crust. The evaluation of the macroscopically averaged
effect of this binding, both as a mechanism for static pinning and also as a
source of temperature dependent resistive drag  -- in addition to that
provided by the Jones formula (\ref{6.20}) -- has been the subject of several
published discussions\cite{AAPS84a}\cite{AAPS84b} \cite{Jones91}\cite{R91} but
there still seems to be considerable disagreement about the quantitative
conclusions to be drawn\cite{Jones93}.  In view of the lack of consensus among
the experts, and because it does not seem to us that any of the discussions we
have cited is entirely satisfactory, what we propose here is a new formula
providing a lowest order approximation of the kind that seems most plausible
to us.

What is generally agreed is that the magnitude $\calF$ say of the force
exerted by an ion on a vortex core with which it is in contact will be
given by 
 \be \calF\approx\EPin/\xiN \eqn{7.10}\fe 
where $\xiN$ is the relevant correlation length characterising the
vortex core radius.  However an essential point that has not always
been emphasised as much as we think it ought to be is that if both the
vortex cores and the ionic lattice were infinitely rigid there would be
{\it no net pinning at all} because the diversely directed forces due
to individual ions would cancel out when averaged over a sufficiently
long length of vortex core.  To legitimately apply the standard formula
giving the pinning force  $\FPin$ per unit length of vortex in the
standard form\cite{R91} by the formula
 \be \FPin\simeq \calF/\bPin \, ,\eqn{7.11}\fe 
where $\bPin$ is the mean distance between  ``pinning
nuclei'' along the vortex, one should be very careful about how this
quantity $\bPin$ is defined.  In the recent discussion by Ruderman\cite{R91} it
is supposed that $\bPin$ is determined by the number of ions that would
happen to fall within  a randomly located straight and narrow tube with
radius equal to that of the vortex core which is of order $\xiN$, so
that one simply obtains  $\bPin\simeq\aC^3/\pi\xi^2$. What we wish to
argue however is that such a purely geometric formula is quite
inappropriate, because the ionic nuclei that are counted thereon will
be uniformly distributed over the vortex core so that the radially
directed forces to which they give rise will entirely cancel out. The
only ionic nuclei that can contribute to the net pinning effect are the
extra nuclei that are relatively displaced by the local pinning force
by a small distance, $x$ say, in the range 
 \be \xiN\lta x \lta\aC \, \eqn{7.21}\fe 
so that they are effectively in contact with the vortex core but would
have been out of range had there been no relative displacement.  It is
these extra nuclei that break the symmetry between forward directed and
backward directed forces, because it is exclusively these extra nuclei that 
will be coherently positioned in such a way that their force contributions can
add up constructively in the same direction.  The restriction
(\ref{7.21}) is equivalent to a restriction on the magnitude of the
magnitude of force per nucleus that is compatible with effective
pinning:  in cases for which the force $\calF$ is too weak, so as to
produce a very small average displacement, $x\lta\xiN$, there will be no
net pinning because only symmetrically distributed nuclei will be
involved; on the other hand if the force $\calF$ is too strong, so that
it would give $x \gta a$, the crystal structure will be overridden so the
vortices will be able to move through the crust as if it were fluid.  

If the individual force per nucleus (\ref{7.10}) actually is in the range
compatible with (\ref{7.21}), then the mean distance $\bPin$ between the
nuclei that effectively contribute to the resulting average pinning
pinning will be roughly expressible in terms of the mean displacement
$x$ by
 \be \bPin\approx {\aC^3\over \xiN x} \, .\eqn{7.12}\fe 
Under such conditions, there actually will be an effective pinning force per 
unit length, which will be given roughly by
\be\FPin\approx {\EPin x\over \aC^3}\, . \eqn{7.13}\fe

As well as the value of $\EPin$ and $\xiN$ -- whose estimation as
discussed in Subsection \ref{6_2}  seems to be the subject of fairly
general agreement \cite{R91}, at least at the level of qualitative
principles, even though the results remain quantitavely rather vague --
all that we still need to be able to apply the formula  (\ref{7.13}) is the
appropriate value of the mean displacement $x$, which does not seem to
have been considered in the articles we have cited, but for which it is
not too difficult to make a rough guess.  There are in principle two
distinct ways in which the relative displacement $x$ can be produced:
either  the vortex core can be bent towards the position of the
attracting nucleus or the ionic nucleus can be pulled aside from its usual
position in the crystal lattice.

Let us first consider the case of vortex core bending. When a segment of the
length of the separation  $\bPin$ between neighbouring pinning positions is
subject to a small lateral displacement $x$, the corresponding bending angle
will be of order $2x/\bPin$ and hence the total of the corresponding force
that must be exerted by the relevant pinning nucleus on the vortex segments on
each side will be given in order of magnitude by
\be \calF\approx{4xT\over\bPin}\, ,\eqn{7.15}\fe
where $T$ is the relevant tension. The value of $T$ that is relevant in this
case will be small compared with the formula (\ref{7.3})  since that
formula took account of all the stress within the comparitively large radius
determined by the intervortex separation, whereas for our present purpose the
only contribution is from  the stress within a comparitively small radius. The
relevant radius will be of the order of magnitude of the segment length
$\bPin$, but its precise value is unimportant because it only comes in via the
logarithmic factor in the relevant analogue of (\ref{7.3}) which will have the
form
 \be T\simeq{\pi\hbar^2\nN\over \muN}\, {\rm ln}\Big\{ {\bPin\over\xiN}
 \Big\} \, ,\eqn{7.16} \fe
so that as a rough order of magnitude estimate, taking the logarithmic factor
to be of the order of unity and using the neutron mass $\mN$ as an estimate
for the value of the relevant effective mass $\muN$, we obtain 
\be T\approx{\pi^2\xiN\EPin\over\aNucl^2} \eqn{7.17}\fe
where $\EPin$ is given by the pinning energy formula (\ref{6.22}).
Equating the bending force (\ref{7.16}) to the pinning force (\ref{7.10})
it can thus be seen using (\ref{7.12}) that the sustainable displacement
will be given by
 \be x\approx {\aNucl\over 2\pi}\Big({\aC\over\xiN}\big)^{3/2}
\, ,\eqn{7.18}\fe
and hence that the separation between the relevant pinning sites will
be given by
\be \bPin\approx {2\pi\aC^2\over\aNucl}\Big({\xiN\over\aC}\Big)^{1/2}\, .
\eqn{7.19}\fe
Applying this in (\ref{7.11}), we reach the conclusion that the pinning
force per unit length obtainable from vortex core bending will be
given by
 \be \FPin\approx  {\hbar^2\aNucl^3\nN\over2\pi^2\mN\aC^{3/2}\xiN^{5/2}}
\, .\eqn{7.20}\fe

Let us now consider the other mechanism that can provide the relative
displacement $x$, namely that whereby the nuclei are pulled aside from their
normal positions in the ionic lattice. Since the crystal structure is due
simply to electrostatic Coulomb repulsion between neighbouring ions, it is
easy to see that in terms of the lattice constant $\aC$ representing the mean
separation between the ions, the restoring force $\calF$ on an ion that is
subject to a small displacement $x$ will be given in order of magnitude by
 \be \calF\approx{Z^2 e^2 x\over \aC^3} \eqn{7.23} \fe
where $e$ is the electronic charge coupling constant and $Z$ is the
ionic charge number, which is expected to be typically of order
$Z\approx{_1\over^2}\times 10^3$.
Equating this to the pinning force (\ref{7.10}) we obtain
\be x\approx{\aC^3 \EPin\over Z^2 e^2\xiN}\, ,\eqn{7.24}\fe
which corresponds to a separation between the relevant pinning sites 
given simply by
\be \bPin\approx {Z^2 e^2\over \EPin}\, .\eqn{7.25}\fe
It follows from (\ref{7.11}) or (\ref{7.13}) that the pinning force per unit length 
obtainable by displacing ions from their lattice sites will be given by
\be \FPin\approx  {\EPin^2\over Z^2 e^2\xiN} \, .\eqn{7.26} \fe

In view of their sensitivity to the rather vaguely known correlation
length $\xiN$, the quantitative evaluation of the pinning forces predicted
by the formulae (\ref{7.20})
and (\ref{7.26}) is not easy, but it is clear that the latter will be
relatively negligible, i.e. the vortex bending contribution (\ref{7.20})
will dominate, except perhaps in the deepest part of the crust
where the two kinds of contribution may be comparable if
$\xiN$ is sufficiently small. It is also clear that even the bending
contribution (\ref{7.20}) will be much too weak to resist the
Joukowski-Magnus force given by (\ref{6.12}) -- i.e. the ratio
\be {\FMagnus\over\FPin}
\approx  { 2\pi^3\mN\aC^{3/2}\xiN^{5/2} v_{_{\rm nv}}\over \hbar^2\aNucl^3 }
\, ,\eqn{7.27}\fe
will be much greater than unity --  whenever the relative flow speed
$v_{_{\rm nv}}$ exceeds a depth dependent critical value corresponding
to a differential rotation rate that in most parts of the crust will be
a small fraction of a revolution per second.

\section{Conclusions}

The new transfusive kind of relativistic superfluid model developed
here in Section \ref{Sec2} can in principle provide the framework for a rough
but realistic representation of the bulk motion of the material in a neutron
star. Such a representation should be useful as a first approximation that can
be taken as a basis on which a more accurate  description including details of
diverse secondary phenomena (notably magnetic effects) can then be developed
by successive approximations. However in order to carry out such a program in
practice it will be necessary to obtain more complete information about the
requisite equations of state and in particular the parameter dependence on the
various forces involved, for which results quoted in Section \ref{Sec6}, as
well as the new formulae derived here in Section \ref{Sec7}, should be
considered just as tentative provisional estimates. These estimates may need
substantial revision to allow for detailed effects that have not yet been
taken into account here at all, and that have not yet been sufficiently
explored in preceeding published litterature. Further work will also required
for the evaluation of the dissipative transfusion rate coefficient $\chemr$ in
(\ref{3.11}), though this is not of primary importance because it can be
expected to be so high that non-dissipative transfusive equilibrium equation
(\ref{2.31b}) should provide an approximation that will be more than
sufficiently acurate for the purposes we have in mind (since the electro-weak
interactions involved will presumably be very rapid compared with the relevant
neutron star evolution timescales). An example of a potentially more important
effect that has not been discussed here is thermal barrier penetration, whose
relevance for pinning has been noticed in previous
work\cite{AAPS84a}\cite{AAPS84b}\cite{Jones91}, but for which it much still
needs to be done before the results can be considered reliable.

Assuming the modifications due to such provisionally neglected thermal
and other effects are not large enough to invalidate the prediction
based on the Jones formula (\ref{6.24}) of a very low range of values
for the drag ratio $\drac$ in the crust, it is to be anticipated that
the resistive damping timescale $\tau$ given by (\ref{6.18}) can be
large enough compared with the pulsar slow down rate $\dot\OmC$ to
allow the build up of a differential rotation with an order of
magnitude $\tau\dot\OmC$ that may easily exceed the critical value
beyond which vortex pinning will break down, which according to the
reasonning in the preceeding subsection requires a difference of a few
revolutions per second at the very most. Beyond this threshold the
vortex lines in the crust will tend to corotate with the neutron
superfluid rather than the -- in general more slowly rotating --
``normal'' matter. On the other hand, in the very high density layers
below the crust the resistive drag is expected to be so high that 
the vortices will in fact be dragged along with the ``normal'' material
in a manner that simulates the effect of pinning, but with the
important difference that this strong drag effect is not subject to a
threshold such as that beyond which pinning in the strict sense will
break down.

\bigskip
{\bf Acknowledgements.}
\medskip

This collaboration has been supported by the C.N.R.S. program ``Jumelage
France - Arm\'enie". The authors also wish to thank S. Bonazzola, E.
Gourghoulon and P. Haensel for instructive discussions.

\end{document}